\documentclass{elsart}
\usepackage{graphicx}

\usepackage{amssymb}

\begin{document}

\begin{frontmatter}

\title{Spectral Mixture Decomposition by Least Dependent Component Analysis}
\author[fzj]{Sergey A. Astakhov\corauthref{cor}},
\corauth[cor]{Corresponding author: Tel.: +49 2461 612526; Fax: +49
2461 612430.} \ead{s.astakhov@fz-juelich.de}
\author[fzj]{Harald St\"ogbauer},
\author[fzj,caltech]{Alexander Kraskov},
\author[fzj]{Peter Grassberger}

\address[fzj]{John von Neumann Institute for Computing,
              Forschungszentrum J\"ulich, D-52425, J\"ulich, Germany}
\address[caltech]{Division of Biology, California Institute of
Technology, Pasadena, CA 91125, USA}

\begin{abstract}
A recently proposed mutual information based algorithm for decomposing data
into least dependent components (MILCA) is applied to spectral analysis, namely
to blind recovery of concentrations and pure spectra from their linear
mixtures. The algorithm is based on precise estimates of mutual information
between measured spectra, which allows to assess and make use of actual
statistical dependencies between them. We show that linear filtering performed
by taking second derivatives effectively reduces the dependencies caused by
overlapping spectral bands and, thereby, assists resolving pure spectra. In
combination with second derivative preprocessing and alternating least squares
postprocessing, MILCA shows decomposition performance comparable with or
superior to specialized chemometrics algorithms. The results are illustrated on
a number of simulated and experimental (infrared and Raman) mixture problems,
including spectroscopy of complex biological materials.
\end{abstract}

\end{frontmatter}


\section{Introduction}
\label{intro}

The problem of estimating parameters (concentrations and pure
spectra) of a linear mixture model underlying a set of spectroscopic
measurements has spurred growth of more than 20 algorithms
\cite{gemperline} that now form an arsenal known as multivariate
self-modeling curve resolution tools (for recent reviews see
\cite{black,brown,geladi,tauler,hopke,lavine,jiang}). Although
advanced to the point that they have been implemented as commercial
software (e.g., \cite{simplisma,soft,mcrals,soft1}), these
techniques still leave room for new developments \cite{mc,m3,btem}.

In their pioneering work, Lawton and Sylvestre \cite{pioneer}
proposed decomposition of mixed spectral signals into {\it
independent} pure components. In practice, this will not be
feasible, since different chemical species do not necessarily have
completely independent spectra (e.g., because they may contain
identical or similar functional groups). Thus, any residual
statistical dependencies in the recovered sources might either
signal a failure of the method, or reflect the fact that the goal of
achieving independent spectra was inconsistent. Aside from that,
there have been only relatively few separate attempts
\cite{nuz,chen,ladroue0,react,ladroue,ren,chin,scholz,pich,buy,huang,visser,shao,gao,bon}
to use statistical (in)dependence of recovered sources (measured,
e.g., by {\it mutual information} \cite{it}) as a criterion in
multivariate curve resolution. However, a more systematic study of
applicability of these methods in spectral analysis is due.

Viewed in a broader context, the use of these statistical
dependencies constitutes the basis for the rapidly growing field of
Independent Component Analysis (ICA) or, more generally, Blind
Source Separation (BSS)
\cite{ica00,ica0,fica,sobi,icac,ica1,ica2,jade,infomax}. One of the
characteristic features of multivariate spectral curve resolution is
that in most spectroscopic techniques the instrumental signals are
nonnegative. In ICA or BSS, sources are not generally assumed to be
of definite sign, although there are a number of papers in the BSS
literature \cite{nmf1,nmf3,nmf2,cichocki,plumbley,nnfastica} that
deal specifically with nonnegative sources. Clearly, such methods
capable of efficient decomposing nonnegative mixtures can be of
potential use in analytical practice.

The ICA approach rests on the underlying assumption of statistical
independence of original (pure) sources \cite{ica1,ica2}. In such a
case pure signals can be recovered by finding a demixing
transformation that minimizes dependencies between estimates. If the
original independent sources and concentrations are positive, then
also the estimates should be so \cite{cichocki} and nonnegativity
can be used as a test of decomposition. Alternatively, nonnegativity
may serve as a target property in optimization, together with other
properties such as maximal independence (see, for example, the {\it
Nonnegative PCA} algorithm \cite{plumbley} and several chemometrics
techniques that use the nonnegativity constraint
\cite{gemperline,mcrals,m3,btem,pioneer,pmf,ae}).

As we already pointed out, in realistic problems the original
spectral signals are not perfectly independent (``overlapping
bands''). These dependencies originate from similarities in the
chemical structure, e.g., due to the presence of similar functional
groups. Such mixtures are typically hard to decompose. This raises
difficulties in straightforward use of general purpose ICA
algorithms \cite{ae}, which can only be studied by detailed
simulations. Here we present a comparative study based on extensive
statistics and show how the ICA methodology applies to realistic
chemometrics problems.

In spectral analysis, the overlap problem has motivated the
development of specialized curve resolution methods. For instance,
one group of algorithms is based on the idea of ``pure variables''
\cite{pv}. A ``pure variable'' is a wavelength at which only one of
the components contributes. Isolation of pure variables for all
mixture components is then a key to decomposition. This way to avoid
the problem  of overlapping bands was successfully implemented in
several algorithms, e.g., KSFA \cite{ksfa}, SIMPLISMA
\cite{simplisma}, IPCA \cite{ipca} and, more recently, SMAC
\cite{smac}). Also, {\it Band-Target Entropy Minimization} (BTEM
\cite{btem,btem2}) has been proposed which involves an explicit
(made by visual inspection \cite{btem}) choice of spectral features
(target regions) to be retained in the course of constrained
optimization. While efficient and highly flexible, supervised band
selection has the drawbacks of being neither fully automated nor
completely blind \cite{black}.

Since ICA is in most applications confronted with signals that
cannot be linearly decomposed into independent sources, a more
proper name would in such cases be the ``Least dependent Component
Analysis''. Analyzing dependencies between reconstructed sources
should then be an integral part of the method. A technical problem
for such an analysis has been the need for fast, robust, precise and
bias free estimators for mutual information (MI). In a recent paper
we employed a novel MI estimator with notably improved properties
\cite{est} in what we called the {\it Mutual Information based Least
dependent Component Analysis} (MILCA \cite{milca}). MILCA combines
several features: (i) high performance ICA algorithm; (ii) output
reliability tests; (iii) cluster analysis of reconstructed sources
for residual interdependencies; and (iv) joining highly dependent
sources into multi-dimensional sources. It has been extensively
tested and compared to other ICA algorithms on various blind
separation problems. MILCA was found to outperform existing
algorithms based on cruder approximations of mutual information (or
other measures of independence) \cite{milca}. Some of those, e.g.,
FastICA
\cite{nuz,chen,ladroue0,react,ladroue,ren,chin,scholz,pich,buy,fica},
JADE \cite{react,huang,visser,jade}, Infomax \cite{shao,infomax},
SOBI \cite{bon,sobi} had been employed in the earlier applications
of ICA to spectral curve resolution. These results suggest that
further progress in developing high-performance ICA methods may well
determine their place in practical analytical spectroscopy.

In this paper we study the potential of MILCA, assisted by proper
data preprocessing and postprocessing, in model-free blind spectral
curve resolution. Unlike specialized chemometrics techniques, MILCA
does not make use of nonnegativity, nor does it rely on band
selection in any form. One could of course include violations of
nonnegativity in some way as a contribution to the cost function
(which would otherwise be just the mutual information) to be
minimized. However, we will not do this here, partly because it is
not {\it a priori} clear how to weigh these two contributions.
Instead, the present study elaborates on how violations of
nonnegativity correlate with the quality of decomposition, and we
will make direct comparisons with other chemometrics algorithms.

The paper has the following structure. Section \ref{meth} introduces
our approach to computing mutual information. It gives a description
of the MILCA algorithm along with the preprocessing, optional
corrections for residual negativity (postprocessing) and measures of
performance used. In Section \ref{data} the data sets analyzed in
this work are briefly described. Section \ref{res} concentrates on
main results and, finally, conclusions are in Section \ref{con}.

\section{Methods}
\label{meth}

\subsection{Mutual information}
\label{minfo}

For a multivariate continuous random variable $(X_1,X_2,...,X_M)$ with given marginal and
joint densities $\mu_i(x_i)$ and $\mu(x_1,x_2,...,x_M)$, the mutual information is given
by \cite{it,est,milca,mic}

\begin{equation}\label{mi}
    I(X_1,X_2,...,X_M)=\sum_{i=1}^M H(X_i)-H(X_1,X_2,...,X_M),
\end{equation}
\noindent where
\begin{equation}
    H(X_i)=-\int \mu_i \ln \mu_i dx_i
\end{equation}
and
\begin{equation}\label{hxm}
    H(X_1,X_2,...,X_M)=-\int \mu \ln \mu \;dx_1 dx_2 ... dx_M
\end{equation}
are the differential entropies \cite{it}.

The MI is a measure of statistical dependence of the $M$ variables,
which means that it is zero if they are strictly independent (i.e.,
their joint density factorizes, $\mu=\prod_i {\mu}_i$), and it is
positive otherwise. The advantages of MI is that it is sensitive to
all types of dependencies (while, e.g., Pearson's coefficient is
sensitive only to linear correlations) and it has a well defined
information theoretic meaning. Also, due to the {\it grouping
property} \cite{est}, it can be decomposed for any partitioning of
the set $\{X_1,X_2,...,X_M\}$ into dependencies within groups of
$X_i$ and dependencies between these groups, e.g., $I(X_1,X_2,X_3) =
I(X_1,X_2) + I((X_1,X_2),X_3)$. This leads directly to a
conceptually extremely simple method of mutual information based
hierarchical clustering (MIC \cite{mic}).

When the variables $X_i$ represent experimental measurements, information about them is
usually given by a finite number of samples (realizations) $x_i^k,\; k = 1,2, \ldots,
N$, and MI has to be estimated by statistical inference. In our present case, the raw
data will be an $M\times N$ matrix ${\bf X}$ of $M$ spectra ${\bf x}_i$ sampled at $N$
wavelengths (frequencies) ${\nu}^k$ each

\begin{equation}\label{mat}
{\bf X}=\left(%
\begin{array}{cccc}
  x_1^1 & x_1^2 & {\ldots} & x_1^N \\
  x_2^1 & x_2^2 & {\ldots} & x_1^N \\
  {\ldots} & {\ldots} & {\ldots} & {\ldots} \\
  x_M^1 & x_M^2 & {\ldots} & x_M^N \\
\end{array}%
\right).
\end{equation}

Note that we view each spectrum as a random variable $X_i$ ($i = 1,2 \ldots, M$) and the
$N$ spectral values $x_i^k,\; k = 1,2, \ldots, N$ as its realizations. This should be
contrasted to the alternative point of view where each spectral value at a given
frequency ${\nu}^k$ defines a random variable $X_k,\; k = 1,2, \ldots, N$, and $x_i^k,\;
i = 1,2, \ldots, M$ are its realizations. Our task consists in computing (and
subsequently minimizing) the MI estimator $\hat I ({\bf X})$ of the spectral data given
by Eq.~(\ref{mat}). Obviously, the realizations ${\xi}^k = (x_1^k, x_2^k, \ldots,
x_M^k)$ are not independent of each other, but in the following we shall neglect this
and use estimators developed for independent identically distributed realizations.

The estimators given in \cite{est} (which are actually closely
related to the estimator for differential Shannon entropies used in
\cite{radical}) are based on $k$-nearest neighbor statistics. They
have been shown to give rather precise estimates of MI in any
dimension $M$. In particular, they seem to be numerically free of
bias for independent signals (when MI is zero). But even when this
is not the case, their bias and variance seem to be smaller than
those of other estimators \cite{est}.

\subsection{Least dependent component approach}
\label{lca}

As in the basic multivariate curve resolution setting, linear ICA starts out with a
mixture model in the form
\begin{equation}\label{mix}
    {\bf X} = {\bf A} {\bf S},
\end{equation}
where {\bf X} is a $M \times N$ matrix of mixed signals, {\bf S} is
a $K \times N$ matrix of unknown pure components $({\bf s}_1,{\bf
s}_2,...,{\bf s}_K)$, and {\bf A} is an $M \times K$ mixing matrix
(concentrations). The problem is to reconstruct blindly {\bf S} and
{\bf A} from the observations {\bf X}, assuming that the original
sources are as ``independent'' as possible. More precisely, the
demixing transformation ${\bf W}$ (which is an estimate for ${\bf
A}^{-1}$ with the superscript $-1$ denoting matrix inverse or
pseudoinverse) is sought such that it minimizes the mutual
information estimator $\hat I({\bf Y})$ of the estimated components
${\bf Y}={\bf W X}$. The ICA decomposition is defined up to scaling
(intensity ambiguity \cite{amb}) and permutation of sources
\cite{cichocki}. In the simplest version of the MILCA algorithm
which we apply here, the MI between components ${\bf y}_i$ is
computed at equal frequencies ${\nu}^k$ (possible dependencies
between sources at different frequencies can be assessed, e.g.,
using ``delay'' vectors; for details on this technique and also for
a noninstantaneous mixing ansatz see \cite{milca}).

The first step in ICA usually comprises principal component analysis (PCA),
also called (pre)whitening \cite{icac,ica1,pca} which minimizes linear
correlations in the data. The number of sources ($K \le M$) is estimated in PCA
by taking only decorrelated components with highest eigenvalues. Then the
demixing transformation splits into a $K \times M$ prewhitening matrix ${\bf
V}$ and a square rotation matrix ${\bf R}$:

\begin{equation}\label{demix}
    {\bf W} = {\bf R} {\bf V}.
\end{equation}

Thus, the ICA problem reduces to searching the minimum of $\hat I({\bf Y})$ under pure
rotation ${\bf Y}={\bf R} {\bf Z}$ of prewhitened vectors ${\bf Z}={\bf V} {\bf X}$. An
important simplification comes from the fact that rotation matrix can be further
decomposed into ${\bf R}=\prod_{i,j}^K {\bf R}_{i,j}$, where each two-dimensional
rotation ${\bf R}_{i,j}$ acts on $({\bf z}_i, {\bf z}_j)$ and is obtained so that it
minimizes pairwise mutual information $\hat I({\bf y}_i, {\bf y}_j)$. Convergence to the
minimum of $\hat I({\bf Y})$ is achieved once all ${\bf R}_{i,j}$ have been optimized
iteratively (for implementation details we refer to \cite{milca}). Drawing a line to
curve resolution methods, ${\bf R}$ resolves rotational ambiguity \cite{amb} by bringing
the estimates as close to the initial independence assumption as possible.

The use of proper preprocessing of data can significantly improve
performance of curve resolution algorithms. Here we study how MILCA
performs on second derivative (SD) spectral data which are known to
be better suited for spectral analysis of complex mixtures
\cite{chen,bon,der1,windig1} than original (raw) spectra. We will
explain why such preprocessing improves ICA decomposition
performance. Specifically, one proceeds from original mixture
vectors ${\bf X}$ to their second derivatives ${\bf X }''$ with
respect to frequency (wavelength), approximated either by finite
differences

\begin{equation}
  \left.{d^{2} x(\nu) \over d{\nu}^2} \right|_{{\nu}^k} \sim x
  ({\nu}^{k-1})-
  2 x({\nu}^{k}) +  x({\nu}^{k+1}),
  \label{diff}
\end{equation}

\noindent or by means of smoothing polynomial Savitzky-Golay
differentiation \cite{sg}. Then, PCA and MI minimization can be
performed on ${\bf X }''$, yielding ${\bf W}''$ and ${\bf Y}''$
(here and below double primes indicate quantities estimated using SD
data, whereas superscript $(0)$ will indicate estimates obtained in
the original space). Due to linearity of Eqs.~(\ref{mix}) and
(\ref{diff}), ${\bf W}''$ represents an estimate for demixing matrix
${\bf A}^{-1}$. The estimates for pure spectra (in the original
space) can be recovered by applying the demixing transformation
${\bf W}''$ on the original measured mixture signals
\begin{equation}\label{double}
  {\bf Y}= {\bf W}'' {\bf X}.
\end{equation}

The MILCA estimates for spectra ${\bf Y}$ and concentrations $\tilde
{\bf A}=({\bf W}'')^{-1}$ can be further refined iteratively through
an alternating least squares (ALS, \cite{mcrals,als}) procedure. In
this method the nonnegativity constraint on spectra and
concentrations is imposed in a postprocessing step (i.e., after
mixture decomposition has been done by some curve resolution
algorithm). Specifically, one can write the ALS iterations as
follows:

\begin{enumerate}
    \item set j=0; initialize $\tilde {\bf Y}_0={\bf Y}$, $\tilde {\bf A}_0=\tilde {\bf
    A}$;
    \item set negative entries of $\tilde {\bf A}_j$ to zero;
    \item update spectra $\tilde {\bf Y}_{j+1}=({\tilde {\bf A}_j}^T {\tilde {\bf A}_j})^{-1} {\tilde {\bf A}_j}^T {\bf
    X}$ using the new (nonnegative) $\tilde {\bf A}_j$;
    \item set negative entries of $\tilde {\bf Y}_{j+1}$ to zero;
    \item update concentrations $\tilde {\bf A}_{j+1}={\bf X}{\tilde {\bf Y}_{j+1}}^T (\tilde
     {\bf Y}_{j+1} {\tilde {\bf Y}_{j+1}}^T)^{-1}$ using the new (nonnegative) $\tilde {\bf Y}_{j+1}$;
    \item $j=j+1$; continue at (2) until convergence ir reached;
\end{enumerate}

We denote the result of ALS iterations by ${\bf Y}^{(a)}$ and ${\bf
W}^{(a)}$. If initial $\tilde {\bf Y}_0$ and $\tilde {\bf A}_0$ are
already close the optimal solution, then the ALS procedure should
give small corrections by eliminating the negativity of ICA
estimates. Notice that the mutual information $\hat I(\tilde {\bf
Y}_j)$ will in general slightly increase during the ALS
postprocessing, although, as we will show below, the decomposition
performance is in most cases considerably improved due to better
fulfilled nonnegativity constraint.

\subsection{Measures of performance}
\label{mea}

Bearing in mind the scaling and permutation ambiguities, a good quality measure
for ICA results is the Amari error index \cite{ica2,milca,radical}, which
quantifies how well the demixing transformation ${\bf W}$ agrees with the true
mixing matrix ${\bf A}$ (if such is known)

\begin{equation}
  P ({\bf W}, {\bf A})= {1\over 2K} \sum_{i,j=1}^K ({|p_{ij}| \over \max_k|p_{ik}|} +
            { |p_{ij}| \over \max_k|p_{kj}|})-1,
  \label{pind}
\end{equation}

\noindent where $p_{ij} = ({\bf W A})_{ij}$. The Amari index $P$ vanishes if ${\bf W}$
deviates from ${\bf A}^{-1}$ only in scaling and permutation of elements, and it
increases as the quality of decomposition becomes poor.

Other measures of decomposition performance are applied to match reconstructed
components with pure original sources. To compare our results with those of \cite{btem},
we shall use the inner product of normalized pure and estimated spectral vectors

\begin{equation}
  i({\bf y},{\bf s})=\frac {({\bf y} \cdot {\bf s})}{|{\bf y}| |{\bf s}|}.
  \label{inn}
\end{equation}

In addition, we introduce a scaled overall measure of positivity of the $K$ vectors
${\bf y}_i$ forming the matrix {\bf Y}
\begin{equation}
  \pi({\bf Y}) = {1 \over K } \sum_{i=1}^K
  \frac {\sum_{j: y_{ij} > 0}  y_{ij}}{\sum_{j=1}^N |y_{ij}|}.
  \label{pos}
\end{equation}

If all vectors are strictly positive, $\pi({\bf Y}) = 1$, while it is less than one
otherwise.

\section{Spectral data}
\label{data}

Four exemplary data sets (A through D) taken from the literature
were chosen to evaluate the performance of MILCA on typical spectral
curve resolution problems. These included both artificial and
experimental mixtures with various number of components $M$, amount
of data points $N$ and quality (e.g., noise level, presence of
experimental background, line widths). Our choice of test problems
was largely dictated by two requirements, which we consider
essential: (i) data are publicly available, were previously used for
the purposes of method evaluation and can be assessed later by
alternative methods; (ii) the number of test problems is large
enough to generate statistics of performance.

\subsection{Simulated 3-component mixtures (A)}
\label{da}

To compose a statistically representative test set of randomized
mixtures we first collected a pool of 99 experimental infrared
absorption spectra in the range $550-3830$ cm$^{-1}$ (822 data
points each) selected from the NIST database \cite{data1}. This set
was designed to contain organic compounds having common structural
groups (halogen-, alkyl-, nitro-substituted benzene derivatives,
phenols, alkanes; alcohols, thiols, amines, esters), so that their
spectra have multiple overlapping bands and are, thereby, mutually
dependent. After that a sample of 10000 random triples of
three-component mixtures ($M=K=3$) was constructed by randomly
choosing normalized pure spectra from the pool and applying random
mixing matrices {\bf A}. The resulting sample represents then a set
of blind source separation problems on strictly positive strongly
dependent sources.

\subsection{Near-infrared data set (B)}
\label{db} The second separation problem was chosen from the
publicly available database \cite{data2} which was established in
\cite{dset} to facilitate evaluation and comparison of chemometrical
methods. The near-infrared ($1100-2500$ nm, 700 data points per
spectrum) test sample first analyzed by Windig and Stephenson
\cite{windig1,windig2} consists of 140 experimental mixtures of five
pure solvents (methylene chloride, 2-butanol, methanol,
dichloropropane, acetone). For this data set both the concentrations
in each mixture and the reference spectra of the pure components are
available. The fractional concentrations of the four mixture
components were chosen form the set
$\{10\%,22.5\%,35\%,47.5\%,60\%\}$. The fractional concentration of
the fifth component were set to make the concentrations add up to
$100\%$ \cite{windig1}.

\subsection{6-component infrared mixtures (C)}
\label{dc} To compare MILCA directly with the BTEM algorithm we also
analyze here the same data as those used in the original work of
Widjaja {\it et al.} \cite{btem} taking 14 randomized experimental
6-component mixtures of toluene, {\it n}-hexane, acetone,
3-phenylpropionaldehyde (aldehyde), 3,3-dimethylbut-1-ene (33DMB),
and dichloromethane (DCM) \cite{footnote}. To test performance, we
used also the reference (pure) spectra of these compounds given in
\cite{btem} (all spectra are FT-IR in the range $950-3200$ cm$^{-1}$
with 5626 intensities). Notice that the authors of \cite{btem}
measured and decomposed a set of 18 spectra (including the
experimental background). For more technical details see the
footnote \cite{footnote}.

\subsection{Raman spectra of brain samples (D)}
\label{dd} True blind source separation by MILCA was performed on the spectral data
measured from human brain samples by Krafft {\it et al.} \cite{krafft}. Taken from
neurosurgical resection material, the normal brain tissues of white and gray matter were
subject to near-infrared Raman microspectroscopy, as were also the tumor specimens of
glioma (astrocytoma $WHO^{\circ}3$) and meningioma ($WHO^{\circ}1$) types. The selected
range was $600-3500$ cm$^{-1}$ covered by 3282 wavelengths. For each of the 4 samples, 20
to 42 measurements had been made resulting in 117 spectra. While variability across the
sample is small, spectroscopically resolved differences between the four distinct
specimens are noticeable. This allows to attempt blind mixture decomposition taking all
117 spectra together to extract common least dependent components in each sample and
estimate their average concentrations. These results obtained blindly by MILCA were
subsequently compared to the same Raman spectra of four pure substances (protein albumin
from bovine serum, lipids from bovine brain extract, cholesterol from lanoline, and
water) which were used in \cite{krafft}, assuming that they approximate the main
constituents of the brain tissues.

\section{Results and discussion}
\label{res}

\begin{figure}
\center
\includegraphics[width=12cm]{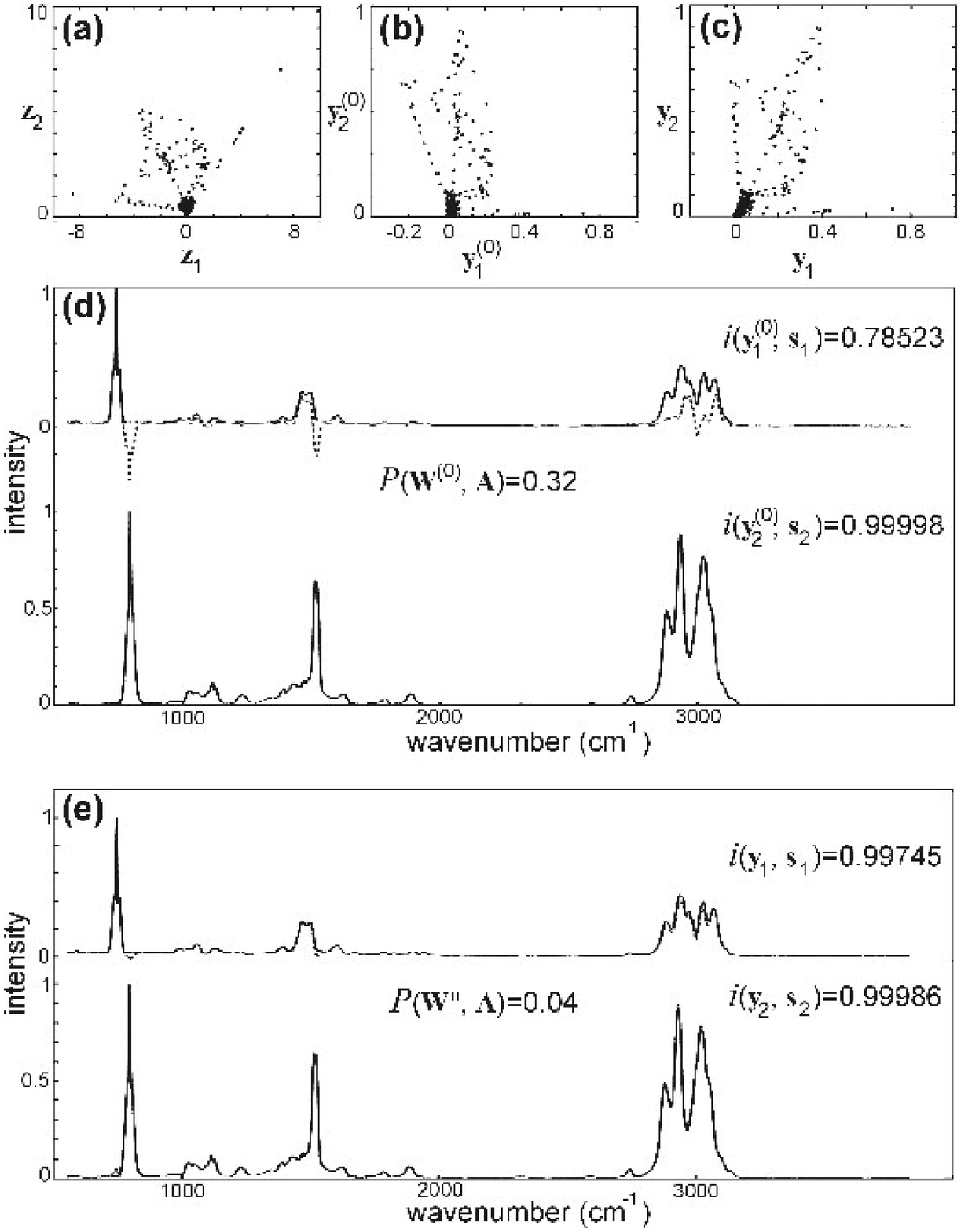}
\caption {PCA and MILCA applied to the mixture of {\it o}-xylene and
{\it p}-xylene. Scatter plots of (a) PCA components $z_1^k$ and
$z_2^k$ ($k=1,2,\ldots,822$) in the original space (${\bf z}_1$ and
${\bf  z}_2$), (b) ICA components ($y_1^{(0),k}$ and $y_1^{(0),k}$)
reconstructed in the original space, (c) recovered components
($y_1^k$ and $ y_2^k$) with ICA done in SD space (see
Eq.~(\ref{double})). Panel (d) shows ${\bf y}_1^{(0)}$ (dashed
curve) and the true pure spectra ${\bf s}_1$ and ${\bf s}_2$ (solid
curves). The component ${\bf y}_2^{(0)}$ is indistinguishable from
${\bf s}_2$. Panel (e) is the same for ${\bf y}_1$ and ${\bf y}_2$
obtained by MILCA in derivative space. In this case, the estimates
are almost indistinguishable from the true sources. Also shown are
the values of the inner product $i$ (Eq.~(\ref{inn})) and the Amari
index $P$ (Eq.~(\ref{pind})).} \label{scatter}
\end{figure}

In order to illustrate potential pitfalls arising from dealing with
highly dependent spectral signals, we first ran MILCA on a simple
synthetic 2-component mixture of {\it o}-xylene and {\it p}-xylene,
two compounds with very similar molecular structures and highly
overlapping spectra (the spectral data were taken from
\cite{data1}). The distribution on Fig.~\ref{scatter}a shows that
prewhitening (PCA) of the initially strictly positive components
already leads to decorrelated vectors that cannot be made positive
by any further pure rotation. Typically, in cases like this,
minimizing the MI results in one component being poorly resolved, in
appreciable violations of positivity and deviations from the pure
signal itself (Figs.~\ref{scatter}b,d). Obviously, this is even more
severe for higher dimensional mixtures. It is therefore likely that
in a problem with originally nonnegative overlapping sources already
the first step (prewhitening) may be counterproductive . This may be
be the reason why algorithms based on PCA followed by subsequent
rotation of decorrelated vectors into the positive domain (e.g.,
\cite{plumbley}) have so far met with limited success only (see also
\cite{ae}). Likely, ICA approaches free of prewhitening might be
more ``compatible'' with such strong constraints as nonnegativity.

When performing ICA in derivative space, the nonnegativity of
components reconstructed by means of Eq.~(\ref{double}) is much
better preserved (see Figs.~\ref{scatter}c,e and figures shown
below). The main reason for the improvements is that preprocessing
with second derivatives removes slowly varying components from the
spectra, and it is these slowly varying components which show most
of the undesired dependencies between pure spectra. Seen from this
point of view, discarding slow components in any multi-resolution
decomposition such as, e.g., a wavelet decomposition \cite{wt} would
presumably have similar effect as taking second derivatives (notice
that we speak here of wavelets in the frequency space, not in the
time domain). This is opposed to selecting nonoverlapping bands
which would correspond to discarding spectral regions. Indeed, one
might apply both techniques (band targeting and filtering out slow
components) in combination, but we shall not do this in the present
paper.

\begin{figure}
\center
\includegraphics[width=10cm]{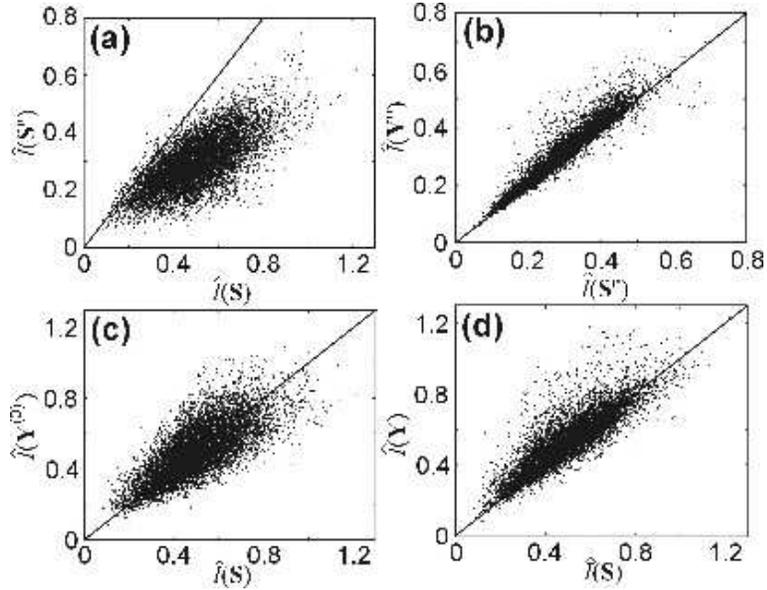}
\caption {Preprocessing and MILCA decomposition of 3-component
mixtures (statistics over 10000 cases, data set A):  (a) MI of the
second derivatives of the sources ${\bf S}''$ plotted against MI of
the original sources ${\bf S}$; (b) MI of estimated SD spectra ${\bf
Y}''$ against MI of the pure SD spectra ${\bf S}''$;  (c) similar to
panel (b), but without using second derivatives at all (MILCA is
performed in the original space, producing estimates ${\bf
Y}^{(0)}$); (d) again similar to panel (c), but with MILCA done
using SD signals, in which case the estimates in the original space
${\bf Y}$ are obtained through Eq.~(\ref{double}). Deviations from
the straight lines indicate differences between quantities plotted
on the vertical and horizontal axes.} \label{stat}
\end{figure}

We next studied behavior of MILCA (with and without
(pre/post)preprocessing) on a large sample of synthetic randomized
3-component mixtures (data set A). To demonstrate the effect of SD
preprocessing, Fig.~\ref{stat}a shows the mutual information of
original triples of sources and those after taking second
derivatives (Eq.~(\ref{diff})). There is a clear trend towards
lowering MI by SD preprocessing. This means that taking derivatives
filters out mostly the information which is common to all 3 spectra,
and which is, therefore, detrimental to MILCA. Since SD acts as a
high-pass filter, the effect can be traced to amplifying rapidly
varying components. The observed gain in decomposition performance
can be further understood by noting that the MI of estimated
components is closer to the MI of pure components, if MILCA is
performed in derivative space (Figs.~\ref{stat}b,c). Finally,
Fig.~\ref{stat}d confirms that the match between MI of reconstructed
signals and MI of original signals (both estimated in the original
space) improves when SD preprocessing is involved (compare to
Fig.~\ref{stat}c).

\begin{figure}
\center
\includegraphics[width=10cm]{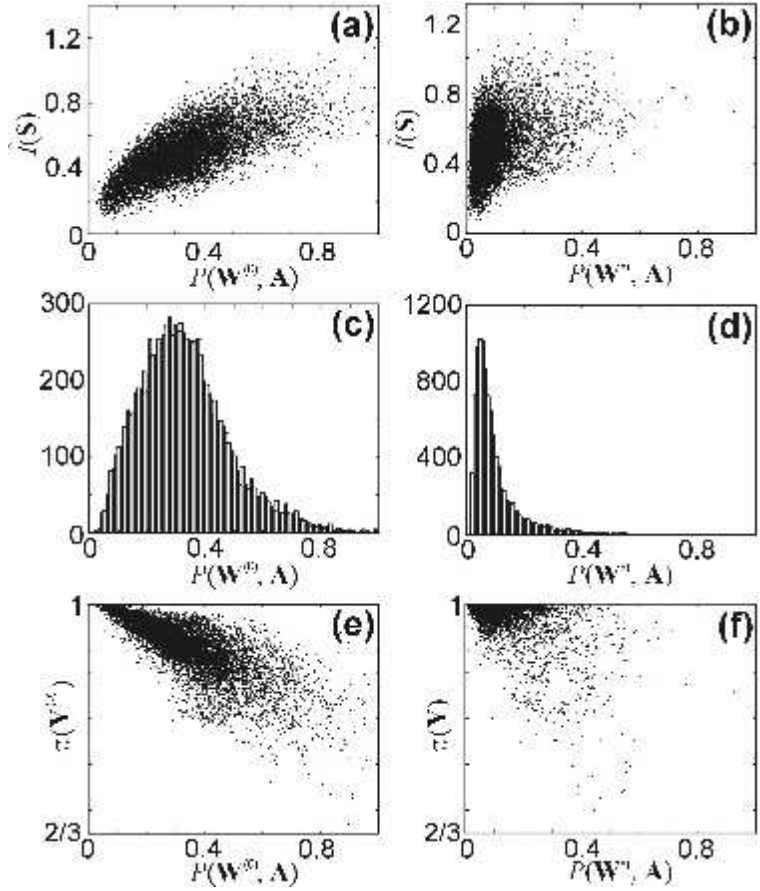}
\caption {MILCA performance statistics (10000 mixtures, data set A). Left (right) panels
show results for MILCA in the original (second derivative) space: (a,b) MI of sources
plotted against the Amari indices $P$ (Eq.~(\ref{pind})) of the decompositions; (c,d)
distributions of Amari indices over the test set; (e,f) dependence of the positivity
measure $\pi$ (Eq.~(\ref{pos})) on the Amari index $P$.} \label{perform}
\end{figure}

A more direct measure of performance (or quality of decomposition) is given by
the degree to which the estimated demixing transformation corresponds to the
actual mixing matrix and how well the nonnegativity is preserved. For this we
gathered the statistics of the Amari index $P$ (Eq.~(\ref{pind})) and
positivity measure $\pi$ (Eq.~(\ref{pos})) over the same test set A. As
expected, the decomposition becomes more difficult as the sources become more
dependent. But while this is very pronounced when MILCA is performed in the
original space (Fig.~\ref{perform}a), it is hardly visible when second
derivatives are used (Fig.~\ref{perform}b). In fact, MILCA with SD
preprocessing was able to reconstruct successfully most of spectra from set A
(Figs.~\ref{perform}d,f), in contrast to MILCA without SD preprocessing
(Figs.~\ref{perform}c,e). Amari index values $P<0.1$ indicate good
decomposition quality, whereas $P > 0.3$ can be considered as unacceptable.
Somewhat surprisingly, the strong (and expected) correlation between Amari
index and nonnegativity seen when MILCA is done without filtering
(Fig.~\ref{perform}e) is nearly completely eliminated with SD preprocessing
(Fig.~\ref{perform}f). Again, this suggests that nonnegativity alone may not be
the optimal target in a PCA-based spectral curve resolution.

In the following applications to experimental mixtures we will only use MILCA with
second derivative preprocessing.

\begin{figure}
\center
\includegraphics[width=12cm]{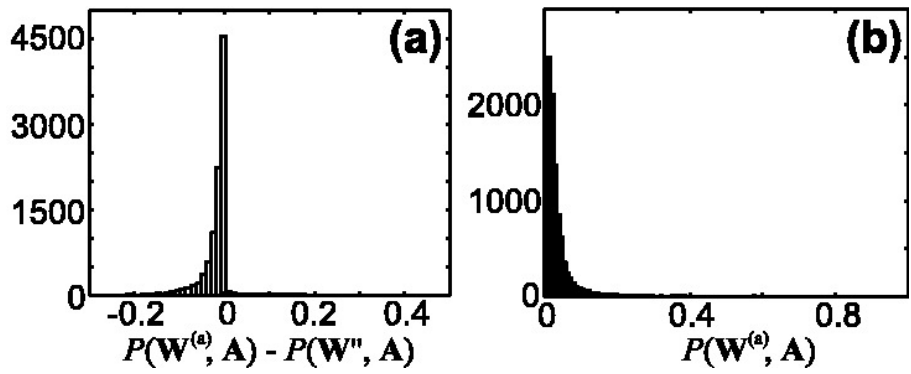}
\caption {Improvement in decomposition performance (a) and the
resulting distribution (b) over the dataset A achieved by applying
the alternating least squares with nonnegativity constraint to the
estimates from MILCA done in the second derivative space (to compare
with Fig.~\ref{perform}d, same scale).} \label{als}
\end{figure}

In order to see if postprocessing might further improve
decomposition performance, we applied ALS with nonnegativity
constraints (as described in Sec.~\ref{lca}, making 600 iterations
for each mixture). Figure~\ref{als}a shows that in the vast majority
of cases the ALS corrections reduced the errors in mixture
resolution and there was only a handful of mixtures (hardly seen in
Fig.~\ref{als}a), for which ALS iterations diverged far from the
MILCA estimates and from the true solutions. The statistics of
decomposition efficiency of the combination second
derivatives-MILCA-ALS is notably improved, with the peak in the
Amari index distribution shifted down to $P \sim 0.05$
(Fig.~\ref{als}b) which is indicative of a reliably high
performance.

We now proceed to experimental mixture problems and comparisons with
several chemometrics algorithms.

\begin{figure}
\center
\includegraphics[width=9cm]{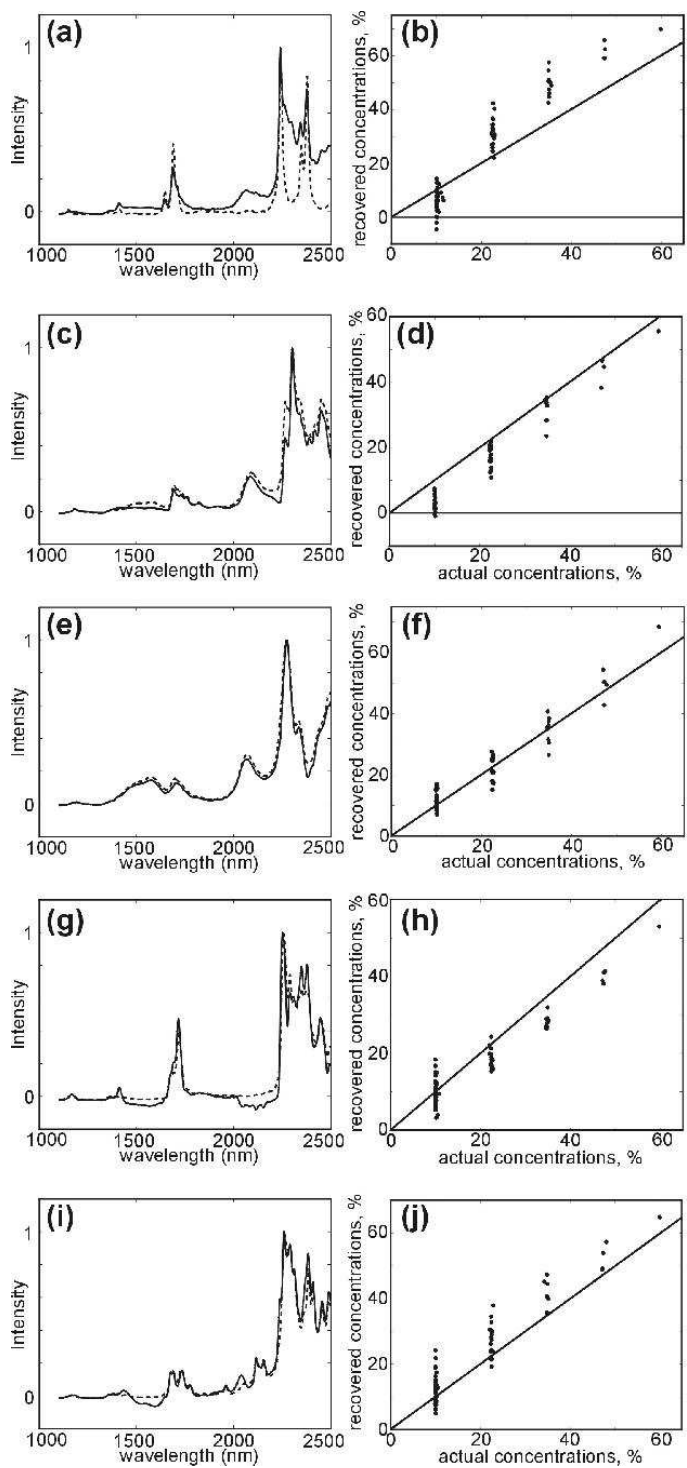}
\caption {Left column: Near-infrared spectra of estimated (solid curves) and original
pure (dashed curves) components (data set B). The components were methylene chloride
(a), 2-butanol (c), methanol (e), dichloropropane (g), and acetone (i). Right column:
Estimated versus actual concentrations of these components. The straight lines have unit
slopes and pass through the origin.} \label{windig}
\end{figure}

The test problem B offered a large set of mixtures to be decomposed,
while each spectrum was relatively scarce in the number of data
points (wavelengths). The five components have strong overlaps in
the most informative range $2000-2500$ nm (see Fig.~\ref{windig}),
with only a few spectral features (also partly overlapping) at lower
wavelengths to facilitate decomposition. Since the signals appeared
rather smooth without much measurement noise, we performed MILCA on
second derivatives computed by finite differences (Eq.~(\ref{diff}))
(we verified that Savitzky-Golay smoothing did not improve results
in this case). The resolved components were found to match the
reference spectra fairly well with only minor violations of
nonnegativity. The largest mismatches are observed for methylene
chloride and 2-butanol (Figs.~\ref{windig}a,c).

To test how well the original concentrations were recovered, in the right panels of
Fig.~\ref{windig} we plot the estimated concentrations versus the original ones. Since
the original concentrations were nearly quantized (see Sec.~\ref{db} and \cite{windig1}),
the vertical scatter of each point cloud indicates the inaccuracies of the source
reconstruction. They are slightly larger than those obtained in \cite{windig1} with
SIMPLISMA, but MILCA produced notably fewer false negative concentrations: they occurred
only for methylene chloride and 2-butanol (Figs.~\ref{windig}b,d) which show also the
worst spectral reconstruction (Figs.~\ref{windig}a,c).

\begin{figure}
\center
\includegraphics[width=7cm]{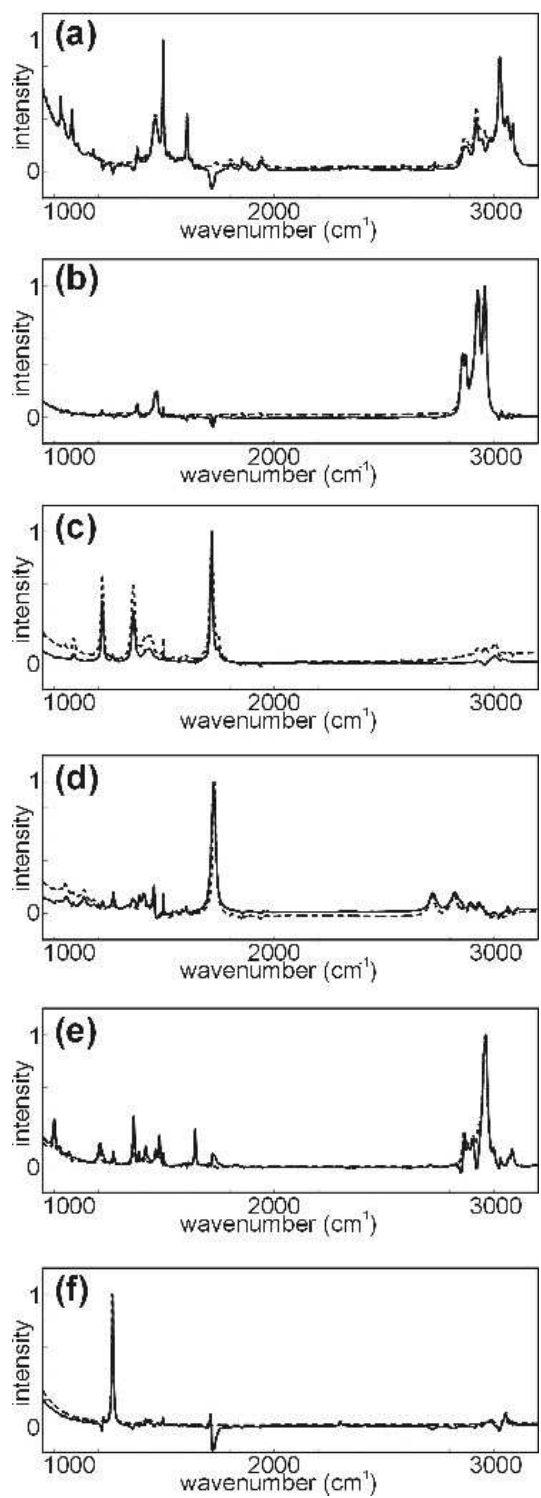}
\caption {Data set C: reconstructed by MILCA (solid) and reference \cite{btem} (dashed)
spectra of toluene (a), {\it n}-hexane (b), acetone (c), aldehyde (d), 33DMB (e), DCM
(f). } \label{btem}
\end{figure}

The next comparison was made with the BTEM algorithm
\cite{btem,btem2}. For this we used the same data (set C) as in
\cite{btem}. The large number of data points per spectrum and
cleanness of the data even without background subtraction
\cite{footnote} allowed to work with second derivatives. In this
case a Savitzky-Golay filter with 81 point window and 7th order
polynomial gave best performance. Figure~\ref{btem} depicts the
resolved components plotted together with the reference (pure)
spectra. Although the MILCA approach does not specifically focus on
preserving certain spectral features, we find all the major bands
reproduced reasonably well. Some distortions appear mostly in the
overlap range $2800-3200$ cm$^{-1}$ which was also a source of
imperfections in the BTEM analysis \cite{btem}. In Table~\ref{tbl}
we give a quantitative comparison between SIMPLISMA, BTEM, and
MILCA, with the inner product $i$ (Eq.~(\ref{inn})) used as
performance measure. Compared to BTEM, MILCA without ALS
demonstrates almost equally high performance while being more
straightforward in not using band selection. However, on this data
set the combination of SIMPLISMA and ALS \cite{simplisma,mcrals}
evidently outperformed both. On the other hand, as reported in
\cite{btem}, the IPCA \cite{ipca} and OPA \cite{opa} algorithms were
less efficient. We find that MILCA+ALS performs clearly better than
BTEM and is close in efficiency to SIMPLISMA+ALS, although much more
statistics would be needed to reliably judge the relative
capabilities of these algorithms.

\begin{figure}
\center
\includegraphics[width=8cm]{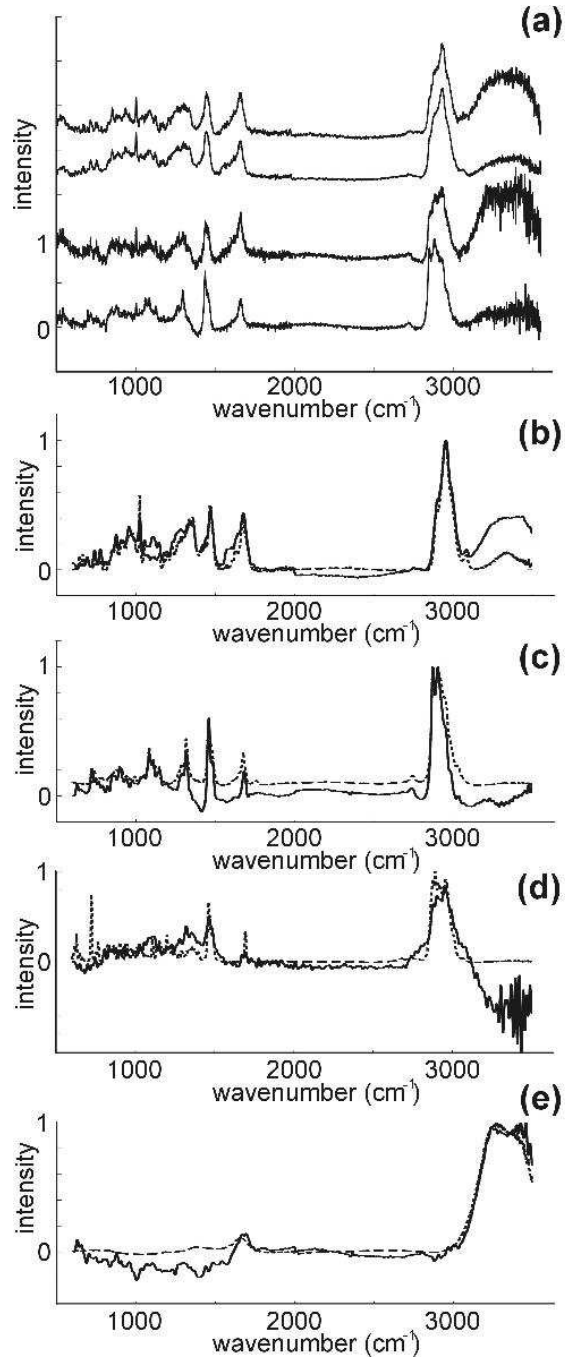}
\caption {Blind determination of composition of human brain tissue
and tumors (data set D). Exemplary near-infrared Raman spectra (a)
of normal white and gray matter, astrocytoma and meningioma tumors
(from bottom to up). The four (b--e) estimated components (solid
curves) plotted to compare with the model set (dashed curves)
\cite{krafft} consisting of protein (b), lipids (c), cholesterol (d)
and water (e).} \label{brain}
\end{figure}

Finally, we proceed to a more realistic, true blind source
separation problem in which no exact information on chemical
composition is known. As noninvasive spectroscopic methods are more
and more used in the analysis of biological materials and {\it in
vivo} measurements
\cite{ladroue0,ladroue,buy,huang,nmf2,krafft,krafft1,krafft2}, they
offer an increasing variety of such ``black'' \cite{black} mixture
separation problems. Here we analyze the results of a Raman
spectroscopy study of brain specimens \cite{krafft} (set D) to see
whether MILCA decomposition could be helpful in quantifying the
abundances of major chemical species present in the brain. In
\cite{krafft} it was shown that a 4-component model can be used to
explain a large part of the complex Raman spectra from brain
samples. In their work Krafft {\it et al.} \cite{krafft} assumed
that the main spectral contributions are that of proteins, lipids,
cholesterol and water, for which reference spectra were obtained.
Then they determined concentrations of these components by making a
linear fit to the experimental data. We attempted to do the same in
a blind manner, applying MILCA to the original experimental set of
spectra. The decomposition was complicated by the high level of
measurement noise (Fig.~\ref{brain}a), so Savitzky-Golay smoothing
derivatives (19 point window, polynomials of order 7) were used to
preprocess the data. The first four least dependent components
resolved by MILCA are plotted in Figs.~\ref{brain}b-e (the fifth and
higher order components contained predominantly noise). We found
that each of them was indeed very similar to one of the spectra from
the model set (dashed curves in Figs.~\ref{brain}b-e), which
supports the result of \cite{krafft} that the model set represents
the main constituents of the tissues. In addition, MILCA gives also
estimates for the mixing matrices, i.e., for the concentrations.
Based on these, we found the following lipid-to-protein average
concentration ratios: 6.5 (white matter), 1.2 (gray matter), 0.5
(astrocytoma), and 0.4 (meningioma). This trend may be of diagnostic
value and is consistent with the model fit parameters of
\cite{krafft} and with the studies by alternative methods
\cite{camp,gai,krafft2}.

\section{Conclusions}
\label{con} We have approached several spectral curve resolution problems by a
new blind source separation algorithm based on accurate estimates of mutual
information (MILCA, available online at \cite{site}). We showed that, with
proper (pre/post)processing, decomposition into least dependent components is
sufficient to achieve separation performance comparable to that of the
state-of-the-art specialized chemometrics techniques.

Least dependent component analysis is a general statistical method
with a very wide range of potential applications (here we refer to
the extensive ICA literature surveyed, e.g., in \cite{ica1,ica2}).
It was designed to perform completely blindly without using any {\it
a priori} or empirical information such as the locations of pure
variables or nonnegativity of original sources. An important
advantage of MILCA over other ICA methods is the fact that it can
use (in)dependencies down to very small scales. Thus it can make
efficient use of high pass filtered signals where most of the
dependencies due to overlapping bands have been reduced. In
practice, this may be helpful, e.g., in cases when reliable
localization of pure variables is complicated by severe overlaps or
noise. Also, opposite to the methods that actually rely on
nonnegativity, MILCA is applicable to alternating sign signals, as
is the case, e.g., in the EPR spectroscopy \cite{ren}.

As our simulations on a large data set (10000 mixtures) have shown,
the use of properly chosen preprocessing (second derivatives) leads
to reduction of nuisance dependencies which would otherwise give
unwanted contributions to mutual information. The latter is the only
cost function in our method. In this sense, filtering out slowly
varying contributions is consistent with the goal of finding least
dependent components in the data performed by minimizing their
mutual information.

Statistics of performance indicates also that imposing nonnegativity constraint
during a postprocessing stage (e.g., in the form of alternating least squares)
can further improve decomposition. We anticipate that the combination of MILCA
and ALS will prove competitive in the category of PCA-based curve resolution
algorithms.

The conceptual simplicity of the MILCA approach is expected to be advantageous
in applied spectroscopy, and this contrasts it with novel but rather
sophisticated curve resolution algorithms \cite{btem,btem2}. On the other hand,
MILCA allows for several generalizations discussed already in \cite{milca}.
These include mixing with small spectral shifts and testing the independence of
spectra at different frequencies. In addition, it should be in principle
feasible to include the nonnegativity constraint directly into the cost
function.

However, more promising are algorithms based on stochastic (Monte
Carlo type) minimization of mutual information subject to the
nonnegativity constraint. The principal advantage of such methods is
that if constrained minimization is performed under affine
transformations (such as, e.g., combinations of rotations and shears
\cite{pmf,ae}), then the prewhitening (PCA) and violations of
nonnegativity it induces can be eliminated altogether. Our
preliminary results (to be reported in a separate publication)
indicate that the decomposition performance of these algorithms may
be superior to their PCA-based counterparts.

\section*{Acknowledgements}
We would like to thank Prof. D.L.~Massart, Prof. P.K.~Hopke  for
pointing us to the useful source \cite{data2} and Dr. W.~Windig for
the data set \cite{windig2}. We appreciate cooperation of Prof.
M.~Garland, Dr. E.~Widjaja who shared their spectra \cite{btem} and
we thank Dr. C.~Krafft for the original data from \cite{krafft} and
discussions. S.A. is grateful to Prof. S.P.~Mushtakova and Dr.
D.A.~Smirnov for suggestions.

\clearpage

\clearpage

\clearpage

\clearpage

\clearpage

\clearpage

\clearpage

\clearpage

\begin{table}
  \centering
  \begin{tabular}{|l|c|c|c|c|c|}

  \hline
   & SIMPLISMA& SIMPLISMA+ALS & BTEM & MILCA & MILCA + ALS\\
  \hline
  toluene  & 0.971 & 0.973 & 0.954 & 0.987 & 0.994\\
  n-hexane & 0.994 & 0.995 & 0.992 & 0.990 & 0.991\\
  acetone  & 0.866 & 0.899 & 0.886 & 0.933 & 0.943\\
  aldehyde & 0.943 & 0.953 & 0.899 & 0.901 & 0.902\\
  33DMB  & 0.576 & 0.963 & 0.983 & 0.964 & 0.948\\
  DCM  & 0.969 & 0.967 & 0.904 & 0.909 & 0.966\\
  \hline
\end{tabular}
  \caption {Data set C: performance of MILCA in comparison to the other
  curve resolution algorithms as measured by the values of inner product,
  Eq.~(\ref{inn}), of reconstructed and reference spectra shown on
  Fig.~\ref{btem}. Numerical data for SIMPLISMA and BTEM were taken from
  Table~3 of \cite{btem}.} \label{tbl}
\end{table}

\end{document}